\documentclass[aps,prl,twocolumn,showpacs]{revtex4}

\usepackage{epsfig}
\usepackage{graphicx}

\DeclareGraphicsExtensions{.eps,.EPS}

\begin{document}

\title{Collective excitations of a dipolar Bose-Einstein condensate}
\author{G. Bismut, B. Pasquiou, E. Mar\'echal, P. Pedri, L. Vernac, O. Gorceix and B. Laburthe-Tolra}
\affiliation{Laboratoire de Physique des Lasers, CNRS UMR 7538,
Universit\'e Paris 13, 99 Avenue J.-B. Cl\'ement, 93430
Villetaneuse, France}
\begin{abstract}

We have measured the effect of dipole-dipole
interactions on the frequency of a collective mode of a
Bose-Einstein condensate. At relatively large numbers of atoms, the
experimental measurements are in good agreement with zero
temperature theoretical predictions based on the Thomas Fermi
approach. Experimental results obtained for the dipolar shift of a
collective mode show a larger dependency to both the trap geometry
and the atom number than the ones obtained when measuring the
modification of the condensate aspect ratio due to dipolar forces. These findings are 
in good agreement with simulations based on a gaussian ansatz.

\end{abstract}


\date{\today}

\maketitle

Interactions strongly affect the properties of quantum degenerate gases \cite{Stringari}. While in most of the experiments to date short range interactions dominate, the recent production of Bose-Einstein condensates (BEC) of highly magnetic Chromium atoms \cite{BECPfau, crBEC}, and the tremendous progresses in the manipulation of heteronuclear molecules \cite{ni}, have brought considerable interest towards particles interacting through long-range, anisotropic, dipole-dipole interactions (DDI) \cite{reviewdipolar}. 

The analysis of collective excitations is an excellent tool to
analyze the effects of interactions in many-body
systems \cite{Stringari,Exp Collective Excitations}. In trapped BECs
where short range isotropic interactions dominate, the excitations at low energies, being
analogous to phonons in homogeneous systems, have a collective character. These
excitations are well understood in the framework of the Gross-Pitaevskii equation (non linear Schr\"odinger equation). DDI add
a non-local character to this non-linear framework. In 3D, it has been predicted that relatively small non-local interactions from dipoles should lead to  modifications of collective modes frequencies. This effect of DDI, which depends on the orientation of the polarization axis of the dipoles with respect to the trap geometry \cite{collectivedip3D,collectivedip3Dbis}, has to date not been observed.  

For larger intensities of DDI in 3D, or in reduced dimensions, the effect of DDI on the properties of BECs is dramatic. In 3D, both the ground state and the collective excitations are strongly modified \cite{bohn}, eventually leading to collapse \cite{Pfau-collapse, santoszoller}. In 2D, the excitation spectrum presents a roton structure reminiscent of the physics of liquid helium \cite{collectivedip2D}. This roton minimum vanishes in 1D, leading to the breakdown of the Landau
criterion for superfluidity \cite{collectivedip1D, collectivedip1Dbis} (similar to what is found with contact interactions \cite{pita-astra}).  Understanding the collective modes of trapped dipolar quantum gases
 has a peculiar interest because both collective modes and long
range interactions are key components in the quantum computing
toolbox \cite{Cirac&Zoller, IonQuantumComputer, zollermolecules}.  

The recent productions of BECs with chromium atoms \cite{BECPfau, crBEC}, which have a relatively large magnetic dipole moment (6 Bohr magnetons), makes it possible to experimentally investigate the effects of DDI on the properties of quantum degenerate gases. Up to now, the only measured effect of DDI on the hydrodynamic properties of BECs has been the modification of the Thomas-Fermi radii \cite{noteTF} of the BEC \cite{stretch}, and its implosion if DDI become comparable to short range interactions \cite{Pfau-collapse}. In this Letter, we report the first measurement of the modification of collective excitations frequencies due to DDI.

In order to experimentally show that a shift of a low energy collective
excitation frequency is induced by DDI, we have measured
the frequency of a collective mode of our spin polarized Cr BEC for
two orthogonal orientations of the magnetic dipoles. The
measured shift is very sensitive to trap geometry, a
consequence of an interplay between the anisotropies of the trap and
of DDI. Despite the fact that the number of atoms in the BEC is
relatively small, the observed shift is in good agreement with zero
temperature theoretical predictions based on the Thomas Fermi (TF) approximation \cite{collectivedip3Dbis}.
By operating at even lower number of atoms, we measure a shift which
decreases, and the deviation from the TF prediction is faster than
what is observed for the aspect ratio of the expanded BEC, in good
agreement with our numerical simulations.

\begin{figure*}
\centering
\includegraphics[width= 5.5 in]{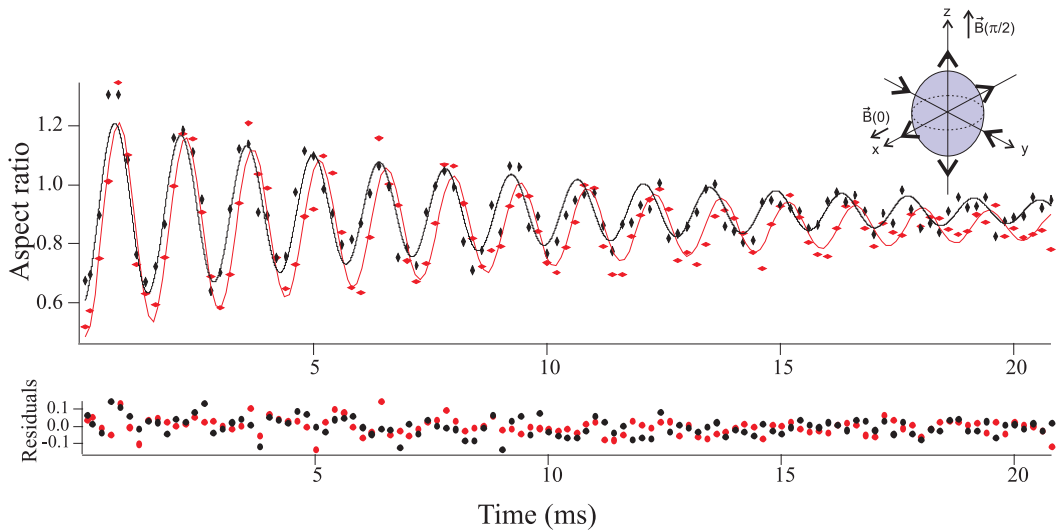}
\caption{\setlength{\baselineskip}{6pt} {\protect\scriptsize
Influence of DDI on the intermediate collective mode frequency. Free
oscillations of the mode after an excitation in a trap set by
$\phi=27 ^{\circ}$ is monitored by measuring the aspect ratio of the
BEC, given by the two experimental TF radii along the y and z axis. The
magnetic field is either vertical (black diamonds), or parallel to
the horizontal imaging beam (red dots). We plot as well the best
fits from damped sinusoidal forms. The residuals to the fit are
shown below. Inset: schematic of the experiment.}}
\label{shiftquadrupole}
\end{figure*}

Our all-optical method to produce Chromium BECs is described in
\cite{crBEC}. Its main specificity is the direct loading of an
optical trap from a Magneto Optical Trap. Atoms are first loaded in
a horizontal (x axis) infra-red (IR) one-beam optical dipole trap
produced by a 1075 nm, 50 W fiber laser. Then some of the IR light
is transferred to a vertical beam (z axis) by rotation of the angle
$\phi$ of a $\lambda/2$ wave plate, to produce a strongly confining
trap in all three directions of space, where evaporation is
performed. The horizontal and vertical beam waists are respectively
equal to 40 and 50 $\mu$m, and $\phi=0$ when all the laser power (30
Watt) is in the horizontal beam. In contrast to \cite{crBEC}, 
the horizontal trapping beam is not retro-reflected, which increases
the stability of the trap: the r.m.s. fluctuations of the BEC TF
radii are reduced from typically 10 percent to below 3 percent.
Condensates can be produced in different trap geometries characterized by oscillation frequencies $\omega_{x,y,z}$, by
adiabatically changing $\phi$ after the quantum regime is reached.

Once the BEC is obtained in a trap for a given value of $\phi$, we
excite shape oscillations of the BEC by modulating the IR laser
power with a 20 $\%$ amplitude during about 15 ms, at a frequency
close to the second surface collective mode resonance (here referred to as the intermediate collective mode). This quadrupole-like mode
corresponds to an oscillation out of phase of the different TF radii
(see fig. \ref{shiftquadrupole}). We then let the cloud oscillate
freely in the crossed optical trap during a variable time. We finally 
switch off the trap, let the BEC expand for 5 ms, and take
destructive absorption images thus measuring the values of the cloud
TF radii. We only report results for the intermediate collective
mode: the lower collective mode is predicted to have a larger shift
due to DDI, but we found that for our trap parameters this mode is
very hard to excite selectively, due to a quasi-degeneracy between
its second harmonic and the higher surface mode (monopole like)
frequency \cite{monoquad}.

A key feature of dipole interactions is its anisotropic nature. In an anisotropic trap, DDI therefore depends on the orientation of the spins (set by a magnetic field) relative to the axis of the trap. We use this property to perform a differential measurement of the shift of the intermediate collective mode due
to DDI: we measure the oscillations of the condensate aspect ratio for two perpendicular orientations of the magnetic field $\theta$. For
$\theta=0$, the B field is horizontal and quasi aligned with both
the imaging beam and the IR trapping beam, while for $\theta=\pi/2$
it is vertical. We then fit the curves by exponentially damped
sinusoidal functions, and deduce the corresponding oscillation frequencies $\omega_Q (\theta)$.
Both the experimental results and the fits are shown in fig
\ref{shiftquadrupole}. We also show the residuals of the fit. The
noise on the experimental data corresponds to a typically 3 $\%$
r.m.s. noise on the measured TF radii. This noise is neither related
to the fluctuations in atomic number, nor to the way collective
excitations are produced. We stress that fig. \ref{shiftquadrupole}
shows that the residuals to the fit do not increase with oscillation
time $t$, a signature of very good short term (20 ms: the typical
oscillation duration) and long term (20 s: the cycling time of the
experiment) stability.

As shown in fig. \ref{shiftquadrupole}, damping of the collective excitations is rather strong, likely due to the large anharmonicity characteristic of an optical dipole trap. The damping rate depends neither on the magnetic field orientation $\theta$ nor on the trap anisotropy set by $\phi$.

\begin{figure}
\centering
\includegraphics[width= 2.8 in]{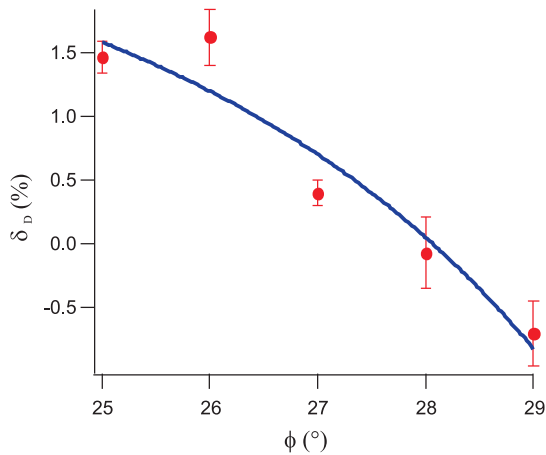}
\caption{\setlength{\baselineskip}{6pt} {\protect\scriptsize}
Variation with the magnetic field orientation of the vertical dipole
mode frequency, as a function of the trap geometry. The experimental
relative shifts $\delta_D$ are plotted versus the angle $\phi$ (see
text). A fit assuming a (constant) tensorial shift and a (trap
geometry dependent) magnetic gradient origin for the shifts is also
shown.} \label{shiftdipole}
\end{figure}

To deduce the shift due to DDI on the collective mode frequency from
the experimental data, it is important to measure and understand all other systematic shifts associated with varying $\theta$. For this, it is in theory necessary to
measure the systematic shift of all three orthogonal vibrational
frequencies of the trap. In practice, the knowledge of the
shift of the vertical axis is the most important, and we
have verified that the effect of the shifts of the
frequency of the other two modes are negligible. We therefore excite
the dipole mode along the vertical direction, and measure the
frequency of the center of mass oscillation of the condensate
$\omega_z(\theta)$. We deduce the relative shift $\delta_D = 2
(\omega_z (0)- \omega_z(\pi/2))/(\omega_z (0) + \omega_z(\pi/2))$,
which is plotted in fig. \ref{shiftdipole}.

Figure \ref{shiftdipole} shows that systematic effects on the
vertical dipole frequency strongly depend on the trap geometry. We
have found two independent sources for this systematic shift. The first is
the presence of magnetic field gradients depending on the applied
magnetic fields. In a parabolic trap, a potential gradient merely
shifts the position of the center of the trap; in a gaussian trap,
in addition to this spatial shift, the oscillation frequency is also
modified. The relative frequency shift is $\delta \omega / \omega=-3
g / m^2 w^2 \omega^4$, where $g$ is the acceleration due to the
potential gradient, $m$ the mass of the atoms, $w$ the waist of the
gaussian potential, and $\omega$ the trap frequency at the bottom of
the trap. As $\omega$ depends on the trap geometry through the angle
$\phi$, this first systematic shift also depends on $\phi$. The
second source of systematic shift, $\Delta$, is related to the
tensorial light shift in Cr: even though the IR laser is very far
detuned compared to the fine structure of the first electronically
excited state, the AC stark-shift of Cr atoms slightly depends on
both the laser polarization orientation, and on the internal Zeeman
state $m$ \cite{chicireanuepjd}. This second systematic shift is
independent of $\phi$. As shown in fig. \ref{shiftdipole},  we fit
our experimental data to a functional form $\Delta -
\alpha/\omega^4$, and deduce the tensorial light shift of chromium
for our experimental parameters $\Delta = 3 \pm 1$ percent, to be compared to
our calculation $\Delta = 1.2$ percent \cite{chicireanuepjd}. To our
knowledge, this represents the first measurement of the tensorial
light shift of Cr.

\begin{figure}
\centering
\includegraphics[width= 2.8 in]{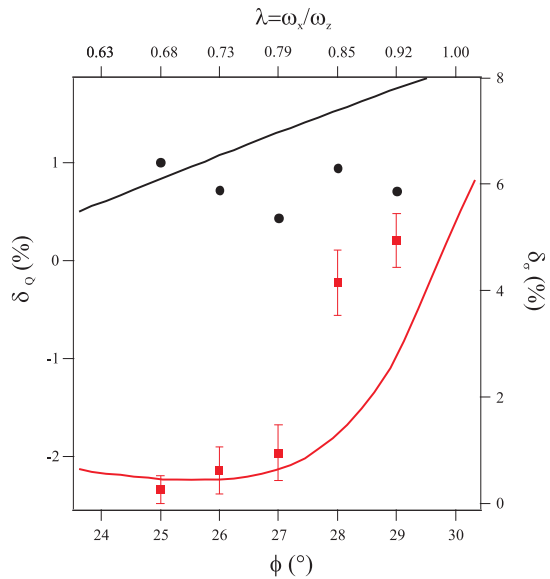}
\caption{\setlength{\baselineskip}{6pt} {\protect\scriptsize }
Effect of the DDI on the experimental relative shifts induced by a
90 degree B field rotation, as a function of the trap geometry:
$\delta_Q$ (red squares) is the shift of the intermediate collective mode
frequency, while $\delta_{\sigma}$ (black dots) is the shift of the
expanded BEC aspect ratio. Results of numerical simulations are also
shown.} \label{comptheoexp}
\end{figure}

Once we have measured the experimental shift of the collective mode
frequency $\delta_{exp} = 2 (\omega_Q (0)-
\omega_Q(\pi/2))/(\omega_Q (0) + \omega_Q(\pi/2))$, we estimate the shift of the collective mode due to
DDI, $\delta_Q$, by subtracting from $\delta_{exp}$ the systematic shift of the intermediate collective mode frequency due to $\delta_{D}$ 
(which we estimate using the theory of \cite{dum}). As shown in fig
\ref{comptheoexp}, we find for various trap geometries, a good
agreement between the experimentally measured $\delta_Q$, and a
numerical model generalizing the theoretical results of
\cite{collectivedip3Dbis} to non axisymmetric parabolic traps in the
TF regime.

We also plot in fig. \ref{comptheoexp} the measured modification of
the aspect ratio due to DDI for the BEC measured 5 ms after its
release from the trap (similar to results reported in
\cite{stretch}). For this, we use the BEC shape oscillation
experimental data (see fig\ref{shiftquadrupole}), and deduce the BEC
equilibrium aspect ratio, $\sigma(\theta)$, from the mean value of
the TF radii. The corresponding relative variation is
$\delta_{\sigma}
=2(\sigma(0)-\sigma(\pi/2))/(\sigma(0)+\sigma(\pi/2))$. We see from
fig. \ref{comptheoexp} that  $\delta_{\sigma}$ is almost constant
for various trap geometries, which is in reasonable agreement with
theory. In contrast, the shift of the collective mode $\delta_Q$
strongly depends on geometry.

The large sensitivity of $\delta_Q$ to trap geometry comes from the
fact that the sign of the shift of the collective mode is
approximately set by the sign of the mean-field due to DDI at the
center of the BEC \cite{collectivedip3Dbis}. As the sign of the
meanfield due to DDI changes (in cylindrical traps) when the trap is
modified from oblate (disk-like) to elongated (cigar-like), the
shift of the collective mode changes sign too. For our
non-axisymmetric trap, we measure $\delta_Q$ in a domain where the
sign of dipole-dipole meanfield flips ($\lambda=\omega_x/\omega_z$ close to 1, see fig. \ref{comptheoexp})), hence the very large
sensitivity illustrated by fig. \ref{comptheoexp}. In contrast, as
DDI always stretch the BEC along the direction set by $\vec B$,
$\delta_{\sigma}$ keeps the same sign whatever the trap geometry is,
hence its lower relative variation with $\phi$.

\begin{figure}
\centering
\includegraphics[width= 2.8 in]{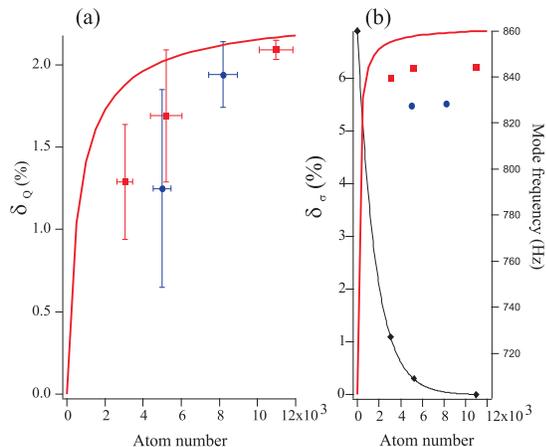}
\caption{\setlength{\baselineskip}{6pt} {\protect\scriptsize }
Variation with the BEC atom number of the relative shifts induced by
a $90 ^{\circ}$ B field rotation, for two different trap geometries (blue circles: $\phi= 27 ^{\circ}$; red squares: $\phi= 26 ^{\circ}$).
(a) variation of $\delta_Q$ and comparison with simulation (for
$\phi= 26 ^{\circ}$). (b) variation of $\delta_{\sigma}$, with the
corresponding results of our simulations. The experimental
intermediate collective mode frequency is also plotted (black
diamonds), the black line is a guide for the eye.}
\label{thomasfermi}
\end{figure}

Although our maximal number of atoms is rather small, our results
for $\delta_Q$ coincide with theoretical predictions for the shift
of collective excitations due to DDI based on the TF approximation.
In our experiment the mean-field due to DDI is only approximately equal to the quantum kinetic energy
$\hbar^2/m R_{TF}^2 \approx 25 $ Hz: the fact that our experimental
results follow predictions based on the TF approximation is
therefore not obvious. To deepen our understanding,
we therefore repeated our measurements for BECs with even lower
numbers of atoms, obtained by loading less atoms in the optical
dipole trap before evaporation. As shown in fig. \ref{thomasfermi}
(a), we observe a rapid decrease of $\delta_Q$ as the number of
atoms is lowered, marking a clear departure from the TF predictions.
On the contrary, fig. \ref{thomasfermi} (b) shows that the shift of
the aspect ratio due to DDI, $\delta_{\sigma}$, is quite insensitive
to the number of atoms. We have checked that for the
variation in number of atoms that we have explored, the collective
frequencies themselves show little shift compared to the TF
predictions without DDI (see fig. \ref{thomasfermi} (b)).

Our results show that measuring collective excitations
is a much more sensitive probe of DDI than measuring the stretching
of the BEC along the axis of the dipoles. We have performed
numerical simulations based on a gaussian ansatz which takes into
account the quantum kinetic energy \cite{youyi}: they confirm that
$\delta_Q$ is more sensitive to a reduction of the number of
particles than $\delta_{\sigma}$ is. Typically, our numerical
results show that it requires about three times more atoms to reach
the TF predictions when measuring shift of collective excitations
compared to when measuring stretching of the condensates. As can be
shown in fig. \ref{thomasfermi} our numerical results are in
relatively good agreement with experimental data. The slight
disagreement between theory and experiment for the collective
excitation frequency shift may indicate the limits of a
simple gaussian ansatz.  

In conclusion, we have characterized the effect of DDI on a
collective mode of a Cr BEC. For large enough number of particles in
the BEC, our results are well explained by TF predictions. In
particular, we find a very large sensitivity of the collective mode
shift as a function of the anisotropy of the trap, a consequence of
the anisotropic character of DDI. We also find that our results
significantly depart from TF predictions for lower numbers of atoms,
even when the striction of the BEC due to DDI is still
very well accounted for by a TF theory. This surprising feature is
another example of the usefulness of collective modes to
characterize quantum degenerate gases. Finally, we have measured for
the first time the tensorial light shift of Cr atoms.

This research was supported by the Minist\`ere de l'Enseignement
Sup\'erieur et de la Recherche (within CPER) and by IFRAF. We thank J. V. Porto for his critical reading of this manuscript.

\end{document}